\documentclass[a4paper]{article}

\usepackage[english]{babel}
\usepackage[utf8x]{inputenc}
\usepackage[T1]{fontenc}
\usepackage{booktabs}
\usepackage{longtable}
\usepackage{footnote}
\usepackage[flushleft]{threeparttable}
\makesavenoteenv{tabular}
\usepackage{array}

\usepackage{chngcntr}
\usepackage{authblk} 

\usepackage[a4paper,top=3cm,bottom=2cm,left=3cm,right=3cm,marginparwidth=1.75cm]{geometry}
\usepackage{adjustbox}
\usepackage{multirow}

\usepackage{amsmath,amssymb}
\usepackage{graphicx}
\usepackage[colorinlistoftodos]{todonotes}
\usepackage[colorlinks=true, allcolors=blue]{hyperref}
\usepackage[resetlabels,labeled]{multibib}
\newcites{S}{References of Appendix}

\title{Supernova Nebular Spectroscopy Suggests a Hybrid Envelope-Stripping Mechanism for Massive Stars}
\author[1]{Qiliang Fang}
\author[1]{Keiichi Maeda}
\author[2,3]{Hanindyo Kuncarayakti}
\author[4,5]{Fengwu Sun}
\author[6]{Avishay Gal-Yam}
\affil[1]{Department of Astronomy, Kyoto University, Kitashirakawa-Oiwake-cho, Sakyo-ku, Kyoto 606-8502, Japan}
\affil[2]{Finnish Centre for Astronomy with ESO (FINCA), FI-20014 University of Turku, Finland}
\affil[3]{Tuorla Observatory, Department of Physics and Astronomy, FI-20014 University of Turku, Finland}
\affil[4]{Steward Observatory, University of Arizona, Tucson, AZ 85721, USA}
\affil[5]{Department of Astronomy, School of Physics, Peking University, Yi He Yuan Lu 5, Haidian District, Beijing 100871, China}
\affil[6]{Department of Particle Physics and Astrophysics, Weizmann Institute of Science, Rehovot 76100, Israel}

\begin{document}
\maketitle


{\bf When nuclear fuel in the core of a massive star with a zero-age main-sequence mass $M_{\rm ZAMS} \gtrsim 8M_\odot$ 
is exhausted, the central part of the iron or magnesium core collapses and forms a neutron star or a black hole $^{\rm \cite{heger03}}$. At the same time, the material above the collapsing core is rapidly ejected, leading to a stripped-envelope supernova (SESN) explosion if the outer hydrogen envelope of the star was removed before its explosion $^{\rm \cite{filippenko97}\cite{galyam17}}$. The envelope is presumably stripped either via strong stellar winds $^{\rm \cite{heger03}\cite{groh13}\cite{smith14}}$ or due to mass transfer to a companion star in a close binary orbit $^{{\rm \cite{eldridge13} \cite{ouchi17} \cite{yoon10}}}$. It is not clear which process is dominant, and whether different mechanisms are at work for different classes of SESNe; type IIb, Ib, and Ic SNe in order of increasing degree of envelope stripping $^{\rm \cite{nomoto95}}$. In this work, a new analysis of late-time nebular spectra of SESNe is presented, which is more sensitive to differences in the core structure $^{\rm \cite{maeda08}}$ than early-phase spectral analysis $^{\rm \cite{liu16}\cite{fremling18}}$. The results show that the progenitors of SNe IIb and Ib are indistinguishable except for the residual amount of hydrogen envelope while the progenitors of SNe Ic are not only deficient in hydrogen and helium, but are also distinctly more massive than SNe IIb and Ib. These findings strongly suggest that more than one mechanism is responsible for the removal of the outer hydrogen envelope and the deeper helium layer, with the former most likely due to binary interaction, and the latter involving a mass-dependent process such as strong stellar winds or episodic pre-explosion mass ejection. }



The final evolution of massive stars leading to an SN explosion, and especially the mass loss mechanism, is an important open problem in modern astrophysics. Without mass-loss, a massive star would form an onion-like layered structure, with a hydrogen envelope, helium layer, and oxygen core from the surface to the inner part $^{\rm \cite{heger03}}$. A massive star explodes as an SN IIb or SN Ib depending on whether its hydrogen envelope is partially or totally removed before its explosion. Further stripping of nearly the entire He layer would result in an SN Ic. SNe IIb can be further divided into extended and compact classes (SNe eIIb and SNe cIIb), where the latter class has a lower amount of residual hydrogen, forming a bridge between SNe IIb and Ib $^{\rm\cite{ouchi17}\cite{yoon17}\cite{cheva10} }$. A special class of SNe Ic (SNe Ic-BL) show broad absorption lines in early spectra, indicating fast expansion velocities and large kinetic energy, and are sometimes accompanied by a gamma-ray burst $^{\rm \cite{iwamoto98}\cite{mazzali02}}$. 


Observations of SESNe pose an apparent conundrum: based on analysis of SN emission during the brightest, opaque phase, it has been argued that SNe IIb/Ib/Ic have similar ejecta mass, 
pointing to similar progenitors for all SESN classes$^{\cite{liu16}\cite{lyman16}\cite{taddia18}}$. On the other hand, the environment in which SNe explode, and in particular the preference of SNe Ic for the most actively star-forming regions (and thus likely more massive progenitors) $^{\rm \cite{anderson12}\cite{kun18}}$ argues otherwise. The analysis of early-phase SN emission is complicated by uncertainties such as opacity and line forming process$^{\rm \cite{dessart12}}$, and the relation between the ejecta mass and $M_{\rm ZAMS}$ is not trivial, as it depends on envelope stripping. Additional progress can be made by identifying a set of observables that can directly trace $M_{\rm ZAMS}$ and the the degree of envelope stripping independently. 

\begin{figure*}[!t]
\includegraphics[width = 15cm]{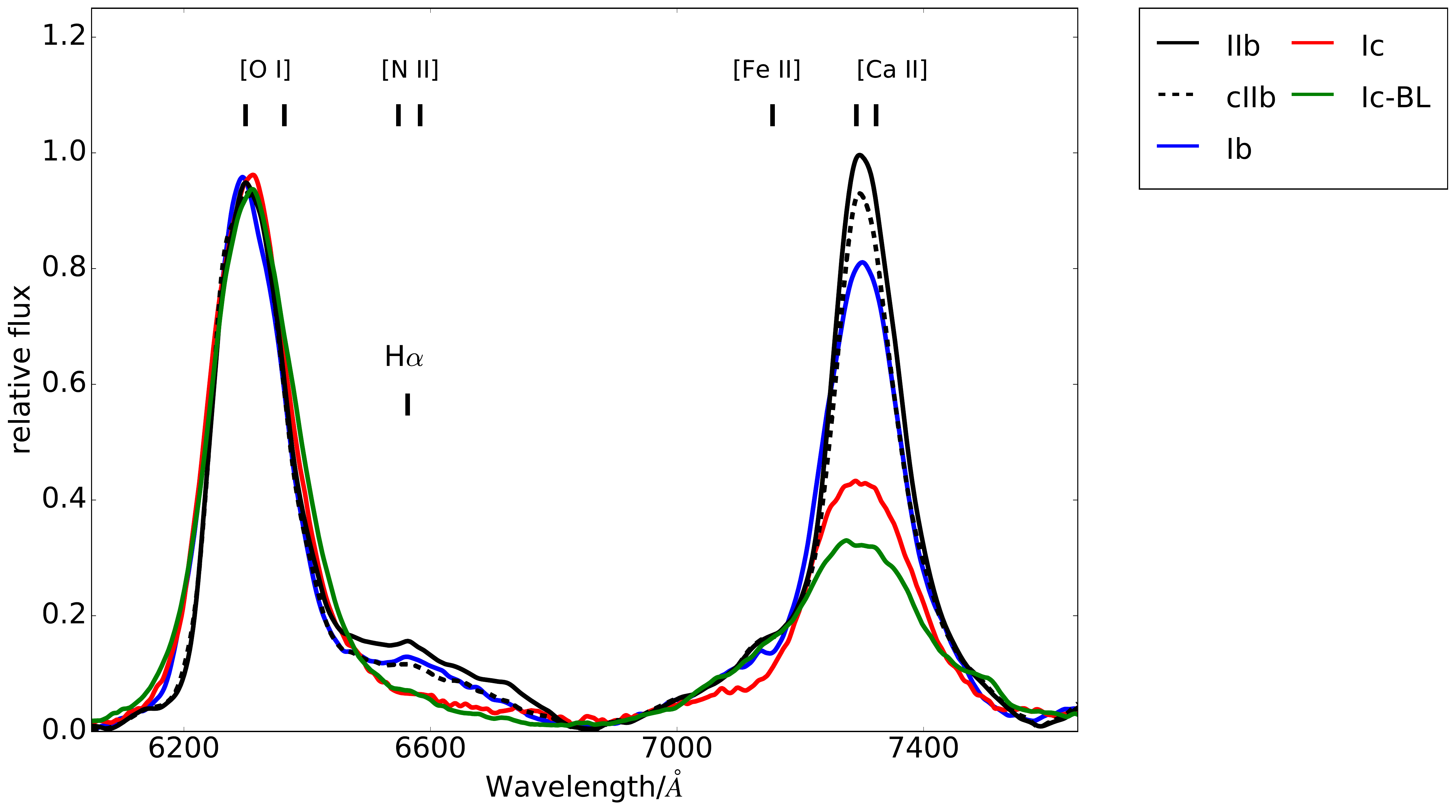}
\centering
\caption{The average spectra of different SESN subtypes are shown. For each object, background emission is subtracted (see Method for details) and the spectra are normalized to the peak of the [O I] doublet. 
}
\label{fig:aver_spec}
\vspace{4mm}
\end{figure*}

In this work, late-time nebular spectra of SESNe IIb/Ib/Ic are collected and analyzed, forming a sample of 12 SNe IIb, 12 SNe Ib, and 22 SNe Ic (see Tables~\ref{tab:IIb_specList} to \ref{tab:IcBL_specList} in Methods). These spectra were obtained around 200 days after maximum light, when SESN have already entered the nebular phase $^{\rm \cite{maeda08}}$. Although there is some small scatter in the spectral phase, it is show in Methods that it will not affect the main conclusion in this work. Figure~\ref{fig:aver_spec} shows the averaged spectra of different subtypes of SESNe. It is clear that the average spectra of SNe Ic and Ic-BL show much larger ratios of the [O I] to [Ca II] doublet. The average spectrum of SNe Ib lies in between those of SNe IIb and SNe Ic. SNe Ib are manifestly different from SNe Ic (and SNe Ic-BL), while the difference with respect to SNe IIb may be statistically insignificant. 

\begin{figure*}[!t]
\includegraphics[width = 15cm]{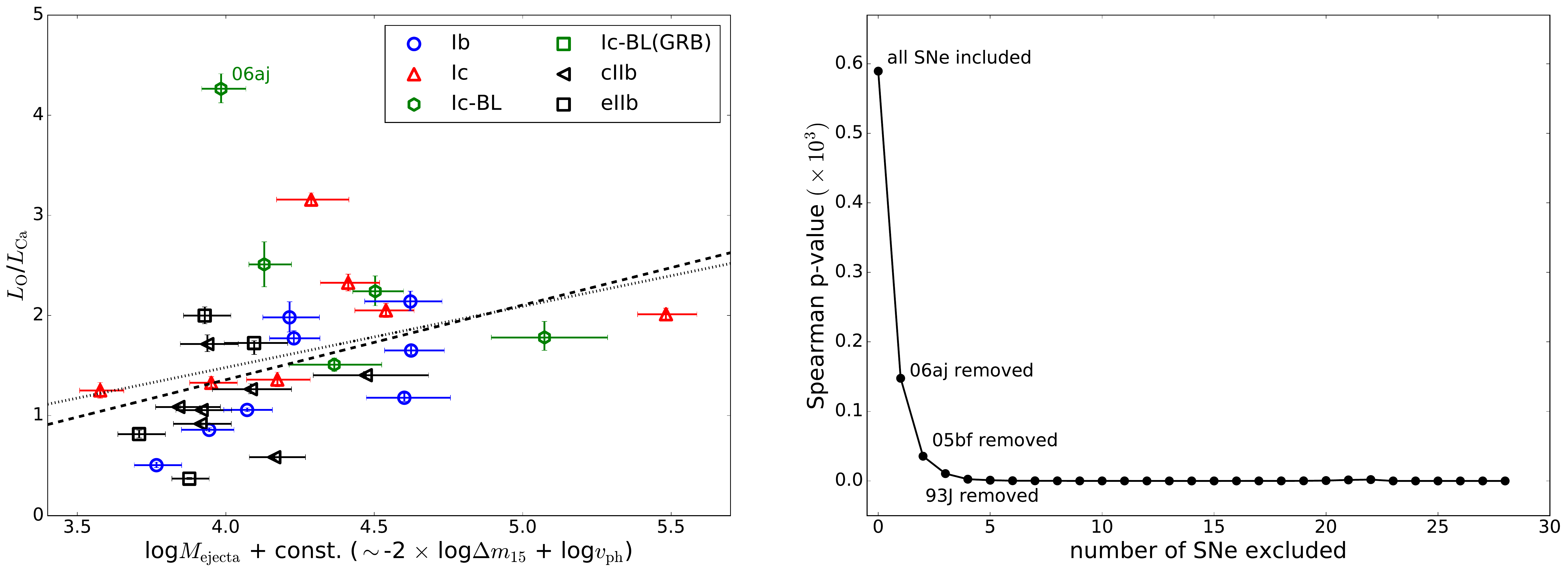}
\centering
\caption{Left panel: The [O I]/[Ca II] ratio is plotted against the measure of the early-phase diffusion time scale (-2$\times \Delta m_{15} + {\rm log}v_{\rm ph}$, where $\Delta m_{15}$ is the magnitude decrease in 15 days from the peak, and $v_{\rm ph}$ is a photospheric velocity). This combination of the early-phase observables characterizes the ejecta mass ($M_{\rm ejecta}$) (see Method). Supernovae of different subtypes are labeled by different colors and symbols. Error bars are given for 1$\sigma$ uncertainties. A positive correlation can be discerned. The dotted line is the best fit to all points, while the dashed line is the best fit when the possible outlier SN 2006aj is excluded. Right panel: To check the possible correlation between the early and late phase quantities further, possible outliers are picked up as follows; One object is removed and the Spearman p value of the residual sample is calculated, and this procedure is repeated until the p value falls below a certain value ($2 \times 10^{-4}$). It is found that SN Ic-BL 2006aj can be regarded as an outlier, and the correlation is significantly strengthened once this is omitted.}
\label{fig:ejecta}
\vspace{4mm}
\end{figure*}

The ratio of [O I] to [Ca II] has been theoretically proposed to be an indicator of the initial progenitor mass ($M_{\rm ZAMS}$). It has been applied to a number of individual SNe $^{\cite{anderson18}\cite{jerk12} \cite{kun15}\cite{maeda07}}$, but has not been tested observationally for a sample of SESNe. Figure~\ref{fig:ejecta} shows how the [OI]/[Ca II] ratio depends on the early phase SN observables, used to characterize their ejecta properties. The horizontal axis traces the ejecta mass via a measure of the diffusion time scale (see Methods for details), and is expected to grow for a more massive star that possesses a larger core. A relatively weak but clear correlation is discerned, which is statistically valid (Spearman coefficient $<2 \times 10^{-4}$ once a prominent single outlier is omitted. See figure caption). While this investigation is limited by the relatively narrow SESN ejecta mass range, it supports the idea that the [O I]/[Ca II] ratio can be used as a measure of $M_{\rm ZAMS}$. It is naturally expected that a significant scatter given that the relation between core mass and ejecta mass is complicated by the existence or absence of the outer envelope, and uncertainties in the conversion of the diffusion time scale to ejecta mass$^{\rm \cite{dessart12}}$. This further motivates the use of nebular spectra as a more direct diagnostic than early-phase data.

A second prominent difference in the average SN subtype spectra is the structure around $\sim 6600$\AA, to the red of the [O I] doublet. An emission feature can be discerned for both SNe Ib and SNe IIb, but it is absent for SNe Ic/Ic-BL. This line has been argued to be dominated by [N II] for SNe Ib and cIIb theoretically $^{\cite{jerk15}}$ and observationally $^{\rm \cite{fang18}}$. This identification is further supported by the similar line flux between SNe Ib and SN cIIb, while SNe eIIb, which likely contain additional hydrogen H$\alpha$ contribution, show stronger excess and drive the total IIb spectrum up when included. This additional hydrogen contribution is most likely related to the more massive H envelope retained in SNe eIIb, but the contribution is minor by day 200 $^{\rm \cite{fang18}}$. Since [N II] is mostly emitted from the outermost region (He-N layer) of the helium envelope $^{\rm \cite{jerk15}}$, this line can be employed as a sensitive tracer of the CNO-cycle processed region and provides a measure of the helium layer stripping, as it can be strong only if the helium layer is almost entirely intact, and should disappear when even a small amount of the helium layer is stripped. 

\begin{figure*}[!t]
\includegraphics[width = 17cm]{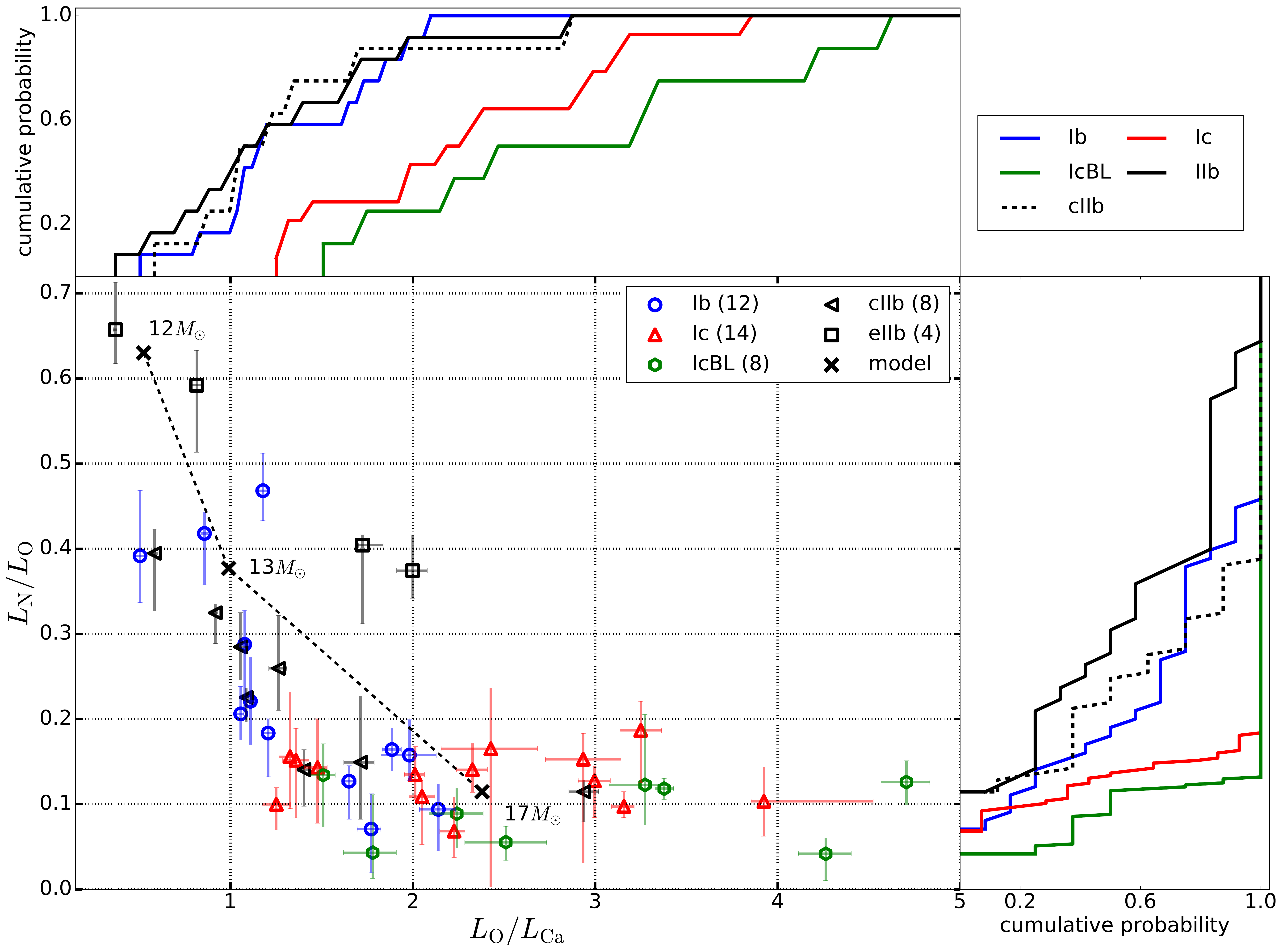}
\centering
\caption {$L_{\rm N}/L_{\rm O}$ ([N II]/[O I]) is plotted against $L_{\rm O}/L_{\rm Ca}$ ([O I]/[Ca II]) in the center panel. The cumulative probabilities of these ratios are shown in the right and upper panels. Objects with different subtypes are labeled by different colors and symbols. Error bars are given for 1$\sigma$ uncertainties. Note that the `[N II]' emission for SNe eIIb deviates from a single relation, which is likely due to contamination by H$\alpha$$^{\cite{fang18}}$. The model predictions for SNe IIb (i.e., no He stripping) are also shown by black crosses, for $M_{\rm ZAMS} = 13, 14$, and $17~M_\odot$ respectively$^{\cite{jerk15}}$. 
}
\label{fig:main}
\vspace{4mm}
\end{figure*}

Figure~\ref{fig:main} shows the distribution of individual SESNe in the [O I]/[Ca II] vs. [N II]/[O I] ratio diagram. The cumulative distribution in the [O I]/[Ca II] ratio shows statistically significant difference between SNe Ic/Ic-BL and SNe IIb/Ib, with only 0.6\% chance probability that SNe Ib and SNe Ic originate from the same population using a Kolmogorov-Smirnov (K-S) test. The hypothesis that SNe IIb and Ib originate from the same population can not be rejected (78.6\% from the KS test) and is consistent with no difference as visually indicated by the cumulative distribution. As the [O I]/[Ca II] ratio is expected to trace $M_{\rm ZAMS}$ irrespective the degree of envelope stripping, this finding provides a strong and direct evidence that the progenitors of SNe Ic/Ic-BL are intrinsically more massive than those of SNe IIb/Ib. On the other hand, the progenitors of SNe eIIb/cIIb/Ib are mutually indistinguishable and can originate from essentially the same progenitors except for the amount of residual hydrogen envelope. 

The same behavior is also evident in the [N II]/[O I] ratio, except for SNe eIIb which likely have a contribution from H$\alpha$ in addition to [N II]. The ratio levels off for SNe Ic and Ic-BL, which may indicate that the He-N layer is totally stripped away for these progenitors. For SNe cIIb and Ib this line ratio is statistically not distinguishable, similarly to the case for the [O I]/[Ca II] ratio. SNe cIIb retain a small amount of hydrogen, and thus the mass stripping probably does not penetrate down to the He layer. While the He layer can in principle be removed to different degrees in SNe Ib, the similarity between SNe cIIb and Ib suggests that SNe Ib suffer from little or no He stripping. The continuous decrease of the nitrogen line ratio toward larger [O I]/[Ca II] ratios can then be interpreted as increased pre-SN He burning for larger $M_{\rm ZAMS}$ stars, which decreases the mass of He-N layer, rather than due to He stripping by any mass-loss mechanism. Given that a large fraction of the helium layer must be stripped away for SNe Ic in order to avoid He line detection in early spectra, and yet there is no distinguishable [N II]/[OI] ratio between SNe cIIb and Ib, He layer stripping apparently does not form a continuous sequence with hydrogen stripping, suggesting there is an intrinsic difference in the stripping mechanism for the hydrogen envelope and the helium layer. 

The model-independent finding in this work can be further strengthened by comparing the observed distribution in Figure~\ref{fig:main} with model predictions for SNe IIb with different $M_{\rm ZAMS}$ which do not include He layer stripping\footnote{https://star.pst.qub.ac.uk/webdav/public/ajerkstrand/Models/Jerkstrand+2015a/} $^{\rm \cite{jerk15}}$. Indeed, the observed distribution closely follows the model prediction. The model shows an increasing ratio of [O  I]/[Ca II] for increasing $M_{\rm ZAMS}$ as expected from the increasing core mass. Taking the model at face value, it is interesting that the [O I]/[Ca II] ratio of SNe IIb/Ib reaches an upper limit of $M_{\rm ZAMS} \sim 17M_\odot$, as this distribution overlaps with the measured progenitor mass distribution of SNe II $^{\rm \cite{smartt09}}$, which do not show substantial H stripping. This may be further evidence that the H stripping process may be independent from $M_{\rm ZAMS}$. SNe Ic arise from more massive progenitors according to the model, with $M_{\rm ZAMS} > 13 M_{\odot}$ and extending to $M_{\rm ZAMS} > 17 M_{\odot}$. The models also roughly reproduce the decreasing sequence in the [N II]/[O I] ratio; in the model this is attributed to more significant He burning for larger $M_{\rm ZAMS}$ (without any He stripping), confirming the observational finding in this work. 

The findings seem to solve the tension regarding SESN progenitors between analysis of early-phase SN emission and SN environmental studies. The results in this work clearly show that the progenitors of SNe Ic are indeed more massive than SNe IIb/Ib, as suggested by  environmental studies. Apparently, previous early-phase ejecta-mass studies
were not sensitive enough to these initial mass differences, and indeed, there is a recent indirect indication from new early-phase SN emission analysis that SN Ic progenitors may be more massive than those of SNe Ib $^{\rm \cite{fremling18}}$. 

The following picture now emerges for the formation of SESN progenitors. The hydrogen envelope of an evolved massive star in a binary is stripped by a mass-insensitive process, i.e., binary mass transfer, producing the Ib and IIb populations that have similar initial mass distributions and no difference in the amount of He layer stripping, as traced by the [N II]/[O I] ratio. However, binary interaction is not efficient enough to further strip the He layer. A less massive star, whose He layer is therefore intact, explodes as an SN Ib or IIb, depending on how much hydrogen is retained. The He layers of more massive SESN progenitors are further stripped by a mass-dependent process, leading to an SN Ic/Ic-BL explosion (that show a statistically significant mass difference with respect to SNe IIb/Ib). This process could be stellar wind stripping$^{\cite{yoon17}}$, or eruptive mass loss toward the end of the life of a massive star, if that occurs selectively for the more massive stars $^{\rm \cite{pasto07} \cite{smith14b}}$. This hybrid picture does not require a large difference in $M_{\rm ZAMS}$ between SNe IIb/Ib and Ic as required for the completely single stellar evolution without binary interaction. Indeed, by comparing the results in this work with model predictions, the characteristic dividing line is likely $M_{\rm ZAMS} \sim 17 M_\odot$. Interestingly, this is the mass range above which SNe II progenitors are not found$^{\rm \cite{smartt09}}$, and the finding in this work may also potentially solve this problem; the mass-dependent mass-loss mechanism may become efficient (only) for the most massive stars, thus effectively reducing the number of H-rich SN II progenitors for this mass range. Further investigation of the mass-dependent process of He layer stripping is now warranted, starting with the outcome of binary interaction models for H-rich envelope stripping which takes place when the progenitor has completed $\sim 90$\% of its life. For stars with $M_{\rm ZAMS} \sim 15-20 M_\odot$, the remaining life time before the SN explosion is $\sim 0.5-1$ million years. Such investigations will fill in the gap between the observed diverse population of massive SN progenitors and theoretical understanding, leading toward a complete understanding of the final stages of massive star evolution.

\clearpage
\newpage

\clearpage
\newpage

\clearpage
\newpage

\clearpage
\newpage

\clearpage
\newpage
\appendix
\counterwithin{figure}{section}  
\section{Sample selection}
\begin{figure*}[!t]
\includegraphics[width = 15cm]{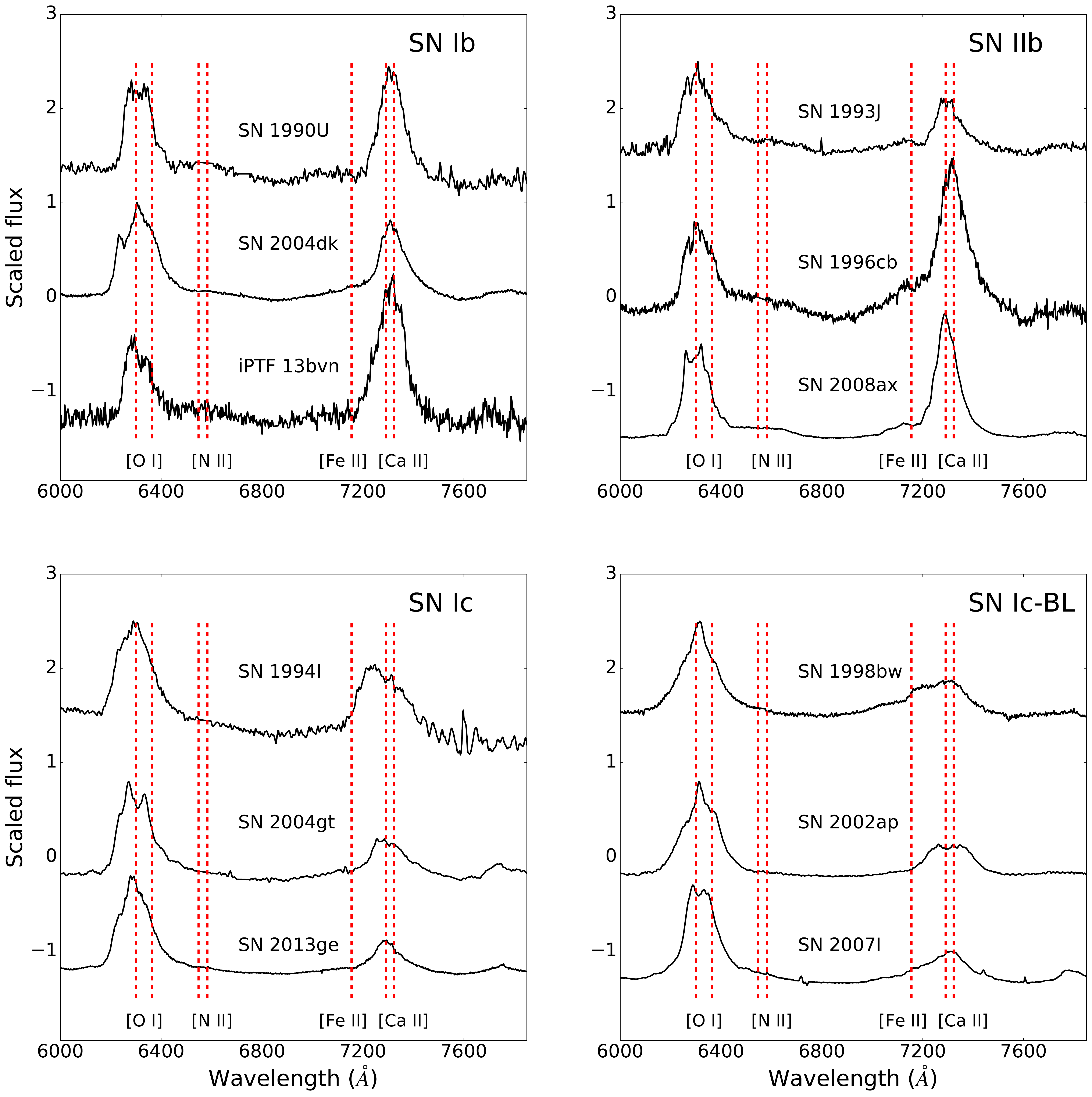}
\centering
\caption{An illustration of SNe of different subtypes in this work. The spectra are corrected for host redshifts and the flux are scaled for demonstration purpose. The red vertical lines are emission lines of interest. From left to right: [O I] $\lambda\lambda$6300, 6363; [N II] $\lambda\lambda$6548, 6583; [Fe II] $\lambda$7155; [Ca II] $\lambda\lambda$7291, 7323.}
\label{fig:spec_illu}
\vspace{4mm}
\end{figure*}

To construct the supernova sample in this work, SESNe (SN IIb/Ib/Ic/Ic-BL) whose nebular spectra (from 150 days $\sim$ 300 days after light curve peak) that are available in the literature are included. A large fraction of the data have been collected at Weizmann Interactive Supernova data REPository (WISeREP)\footnote{https://wiserep.weizmann.ac.il/}$^{, \citeS{yaron12}}$ and The Open Supernova Catalog\footnote{https://sne.space/}$^{, \citeS{guillochon17}}$. The discovery date is restricted to be later than 1990. The sample of this work includes nebular spectra of 46 stripped-envelope supernovae (SESNe), with 12 SNe Ib, 12 SNe IIb, 14 SNe Ic and 8 broad line SNe Ic (SNe Ic-BL). The spectra and their sources are listed in Tables~\ref{tab:IIb_specList} to \ref{tab:IcBL_specList}. The classifications and the phases of the objects are also listed in the same tables. Spectra of some objects in the sample of this work are illustrated in Figure~\ref{fig:spec_illu}.

\section{Data processing}

\subsection{Early light curves}
The explosion parameters are calculated from $V-$band light curves, which are derived from different sources (see Table~\ref{tab:early_para}). The ejecta mass can be estimated from the combination of light curve width and photospheric velocity, using a frequently-adopted scaling relation$^{\cite{lyman16}\citeS{arnett82}\citeS{valenti08}}$: 
\begin{equation}
  w_{lc} \propto \frac{M^{3/4}_{\rm ejecta}}{E^{1/4}_{\rm K}} \ ,
 \label{eq:width_Mej}
\end{equation}
where $w_{lc}$, $M_{\rm ejecta}$ and $E_{\rm K}$ are the width of a light curve, ejecta mass and kinetic energy ($E_{\rm K} \propto \frac{1}{2}M_{\rm ejecta}v_{\rm ph}^2$) respectively. In this work, the width of light curve is estimated from a low-degree polynomial fit to the photometric data around the peak, and is characterized by $w_{lc}\propto\Delta m_{15}^{-1}$, where $\Delta m_{15}$ is the change of $V-$band magnitude in 15 days since the peak. Photospheric velocities of SNe in this work are taken from the literatures. If the photospheric velocity of an SN is not available from the literature, the average value of each subtype is employed (SN IIb: 8300 $\pm$ 750; SN Ib: 9900 $\pm$ 1400; SN Ic: 10400 $\pm$ 1200; SN Ic-BL: 19100 $\pm$ 5000; units: km s$^{-1}$)$^{\cite{lyman16}}$. By taking all these factors together, the ejecta mass can be estimated as: 

\begin{equation}
  {\rm log}M_{\rm ejecta} ~\sim~ -2\times{\rm log}\Delta m_{15}~+~{\rm log}v_{\rm ph} + {\rm constant} \ .
 \label{eq:final_Mej}
\end{equation}

Error in the ejecta mass estimation consists of two parts: error of the light curve width and error of the photospheric velocity. The first part is estimated by shifting peak date within the uncertainty (a typical value is 1 day). The difference is employed as the 1$\sigma$ error of the light curve width. The error of the photospheric velocity is taken from the literature. For objects whose photospheric velocity is not available, the error is estimated to be the velocity dispersion within the subtype to which the object belongs. The estimated widths (characterized by $\Delta m_{15}$) and the sources of the light curves are listed in Table~\ref{tab:early_para}. The photospheric velocities and references are also listed in the same Table.

\subsection{Nebular spectra}
A nebular spectrum is smoothed, de-reddened and corrected for redshift (of the host galaxy taken from HyperLeda\footnote{http://leda.univ-lyon1.fr/}$^{, \citeS{makarov14}}$) before the line decomposition$^{\cite{fang18}}$. The extinction values $E(B - V)$ are derived from the literature (see references in Tables \ref{tab:IIb_specList} to \ref{tab:IcBL_specList}), most of which are estimated from their color evolution. For objects whose multi-band light curves are not available, the extinction is calculated from the equivalent width ($EW$) of Na I D near the light curve peak$^{\citeS{turatto03}}$. For objects for which neither of a light curve nor an early phase spectrum are available (or if the noise of an early phase spectrum is too large for $EW$ of Na I D to be calculated), its $E(B - V)$ is set to be 0.36 mag, which is the average value of SN Ib/Ic$^{\citeS{drout11}}$. The estimation of $E(B - V)$ is quite uncertain, and not always associated with well-determined uncertainties. Although the scatter in the extinction value will affect the absolute scale of the light curves or individual line luminosities, the quantities of interest in this work ($\Delta m_{15}$ and the line ratios) are not sensitive to $E(B - V)$, therefore it is indeed negligible. The background emission is estimated to be the line connecting the two minima at both wings of the emission complex. After the background is subtracted, the emission complex is decomposed.

\begin{figure*}[!t]
\includegraphics[width = 15cm]{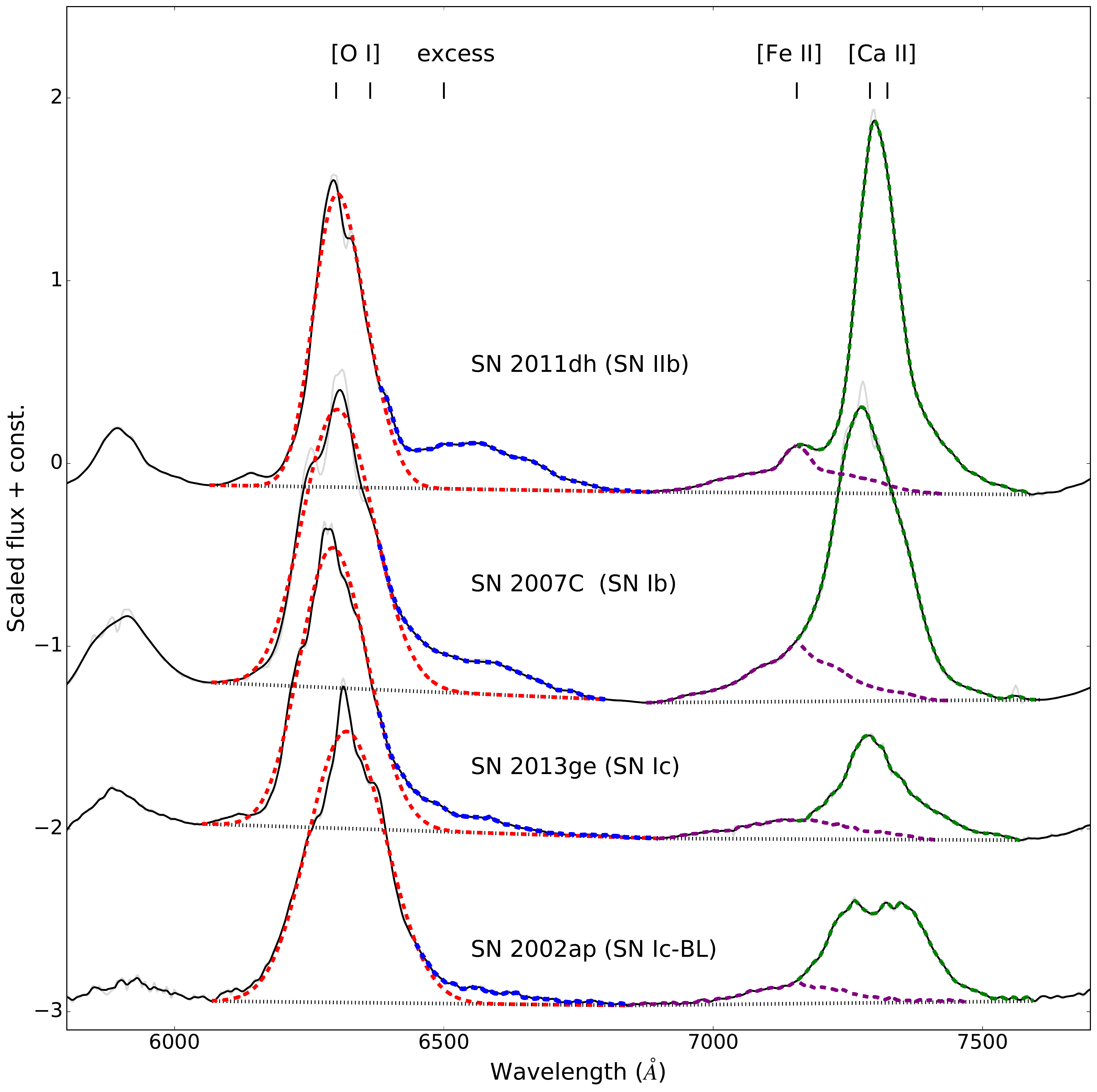}
\centering
\caption{Examples of the line decomposition for SNe of different subtypes. All spectra are corrected for redshift and extinction. The black dotted line is background emission and the red dashed line is the best-fitted double Gaussian profile of the [O I] doublet. The excess flux ([N II]) is illustrated by blue dashed line. Purple dashed line defines the blue wing of [Fe II], together with its mirror image relative to 7155 \AA~. The green dashed line is the [Ca II] doublet.}
\label{fig:line_decom_example}
\vspace{4mm}
\end{figure*}

Examples of the line decomposition of different SN subtypes are illustrated in Figure~\ref{fig:line_decom_example}.
\begin{itemize}
\item $[O ~I]/[N ~II]~complex$: The oxygen line in this complex is fitted by a double Gaussian profile. The Gaussian profiles have the centers located at 6300 and 6363 \AA\ with the same standard deviation, and the flux ratio of the peaks is fixed to be 3:1. The excess flux of this complex relative to the fitted oxygen profile is estimated to be the flux of [N II]. This method will give a none-zero value for the [N II]/[O I] ratio for SNe Ic/Ic-BL. However, this does not necessarily mean that the He-N layer is still present for these objects. Other factors may contribute, including contamination from the correlated star-forming region or continuum scattering from electrons and dust$^{\citeS{jerk17}}$, and therefore the flux may well be leveled off when the [N II]/[O I] ratio is practically zero. 
\item $[Ca~ II]/[Fe~ II]~complex$: It is more difficult to fit the [Ca II] profile in this complex and separate it from other contributions (e.g., [Fe II]) than the case for the [O I] in the [O I]/[N II] complex. Therefore it is not de-composed by a multi-line fitting in this work. Instead, the flux of this complex is estimated phenomenologically from the integrated flux in the wavelength range of interest, which is dominated by [Ca II] with a smaller contribution from [Fe II]. To subtract the flux of the [Fe II] from the complex, the integrated flux of the smoothed spectrum is first calculated. The lower and upper limits in the wavelength range for this integration are set to be the minima at the blue wing of the complex and 7155 \AA\ (i.e., the central wavelength of [Fe II]). Assuming that [Fe II] has a roughly symmetric profile with respect to its central wavelength, the flux of [Fe II] is then estimated by multiplying this integrated flux by a factor of two. This method is based on the assumption that the flux at $\lambda~<$ 7155 \AA\ is not contaminated by [Ca II]. The typical line width of [O I], which is estimated from the double-Gaussian fit, is 50 $\sim$ 60 \AA\, corresponding to 60 $\sim$ 70 \AA\ for [Ca II] if the velocity dispersion of [Ca II] is at the same level. In this case, 7155 \AA\ sits outside the 2$\sigma$ interval, and the contribution of [Ca II] at the bluer wing should be therefore negligible. With the flux of [Fe II] thus determined, this is then subtracted from the total integrated flux of the complex to give the flux of [Ca II]. 
\end{itemize}

\subsection{Error estimation of the line ratios}
Line ratio errors consist of several parts:
\begin{itemize}
\item $Poisson~noise$: The noise level in a nebular spectrum will affect the flux measurement in several ways, including background determination, line fitting, etc. To estimate the error related to the Poisson noise, the degree of smooth is increased (overly smoothed) or decreased, and the deviation of line ratio estimated in this way is associated to the error contributed by the Poisson noise. 
\item $Background~determination$: Although the edge of the [O I]/[N II] complex is well determined for SNe IIb and some SNe Ib, it can be easily contaminated by background for SNe Ib/Ic/Ic-BL with weak or absent [N II] emissions. The uncertainty from the background determination is estimated by shifting the minima, which determine the edges of the complex, within 50 $\sim$ 100 \AA, then re-measuring the line ratios. The difference between the original line ratio and that measured in this way is estimated to be the uncertainty contributed by background determination. 
\item $Line~fitting$: The line profile of [O I] can affect the edge of the [N II] feature, and therefore can affect its flux. The main budget of the uncertainty from the line fitting comes from the uncertainty of the width of [O I]. By changing the width of [O I] (by 1$\sigma$ estimated from line fitting), one can have different estimation of the line ratios. The difference is then employed to be the uncertainty from the line fitting. 
\end{itemize}

The quadratic sum of these errors is estimated to be the total error of the line ratio.

\section{Possible time dependence}
Spectra of different SNe were obtained at different phases, therefore it is important to investigate whether time evolutions of the line ratios could affect the conclusion in this work. The mean phase of the spectra used in this work is $\sim$ 220 days after the light curve peak, with the standard deviation $\sim$ 40 days. The mean phase and the standard deviation of the spectra used for different SNe subtypes are as follows: 207 $\pm$ 38 days (SN Ib), 230 $\pm$ 37 days (SN IIb), 215 $\pm$ 38 days (SN Ic)  and 217 $\pm$ 29 days (SN Ic-BL). No statistical difference between spectral phases among different SNe subtypes can be discerned from a KS test. Since the span in the spectral phase is small, the possible effect of spectral evolution should not be significant, which will be tested in this section.

\begin{figure*}[!t]
\includegraphics[width = 16cm]{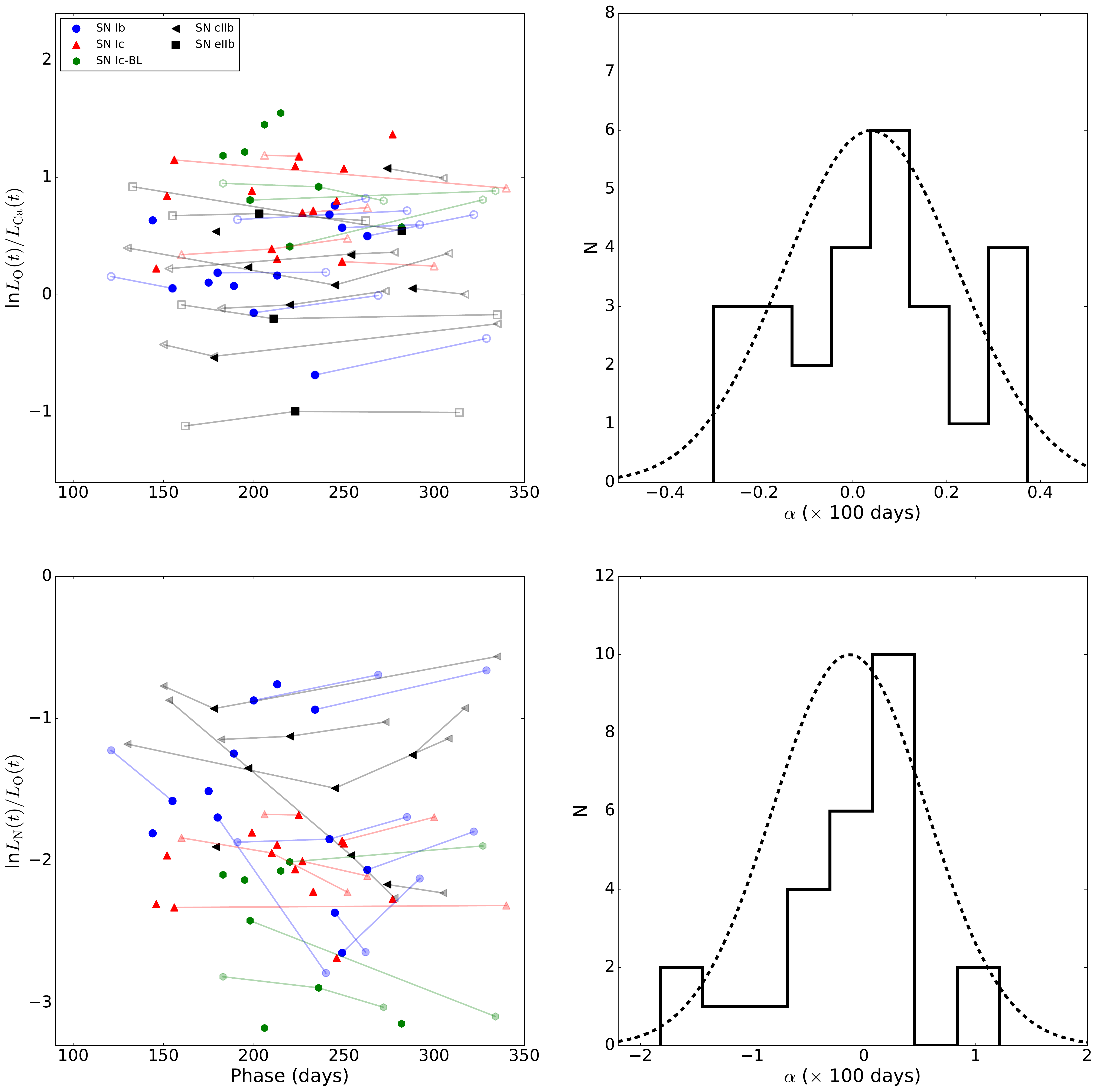}
\centering
\caption{Left panels show evolution of the line ratios in a logarithmic scale. Upper left panel is for [O I]/[Ca II] and lower left panel is for [N II]/[O I]. SNe of different subtypes are labeled by different colors and symbols. The solid points are the line ratios adopted in the main text and the transparent points are line ratios at other spectral phases. Data from the same objects are connected by lines. Since SNe eIIb contain additional contribution from H$\alpha$ in the `[N II] feature', they are omitted in the lower panels. Right panels show distributions in the speed of the evolution for these two line ratios, characterized by $\alpha$ (see text for details). The dashed lines are Gaussian functions (which are adopted in the Monte-Carlo simulations), with the same mean values and standard distributions as the observed distribution.}
\label{fig:line_ratio_evolution}
\vspace{4mm}
\end{figure*}

A direct way to investigate the effect of time evolution is to measure line ratios at different phases for a single object. The luminosities of radioactive-powered metal lines ([N II], [O I] and [Ca II]) evolve quasi-exponentially as a function of time, therefore their line ratios are also expected to evolve in the same manner, $r(t)~\sim~e^{\alpha t}$, where $r(t)$ is the line ratio at phase $t$ and $\alpha$ characterizes the speed of its evolution. In Figure~\ref{fig:line_ratio_evolution}, the line ratio (in a logarithmic scale) is plotted against the spectral phase. The solid points are the line ratios applied in the main text and the transparent points come from spectra in other phases. Points for the same object at different phases are connected by a line. 
It is clear that the [O I]/[Ca II] ratio hardly evolves as compared with the overall scatter, which is consistent with the slowly evolving nature of this ratio reported in the literature$^{\rm \cite{kun15}\citeS{elmhamdi04}}$. The mean values and standard deviations of the slopes in logarithm scale are 0.039 $\pm$ 0.185 /(100 days) for [O I]/[Ca II] (right panel in Figure~\ref{fig:line_ratio_evolution}). On the other hand, the [N II]/[O I] line may evolve relatively quickly for a few objects; The mean values and standard deviations of the slopes in logarithm scale are -0.125 $\pm$ 0.688 /(100 days) for [N II]/[O I]. These slopes correspond to a change in the line ratio by $\sim$ 25\% (for [O I]/[Ca II]) and $\sim$ 60\% (for [N II]/[O I]) in a 100 days time span. 

The above discussion suggests that the evolution of the line ratio may require further consideration. However, except for the intensively-studied SNe IIb and some SN Ic-BL (for example, SNe 1998bw and 2002ap), spectra covering a wide phase range are not available for most objects in this work. The slope calculated from only two points can also be very uncertain. To test how the evolution of the line ratio could affect the conclusion in the main text, 10$^4$ Monte-Carlo simulations are carried out. In each simulation, a randomly-generated slope (which is assumed to be normally distributed, with the mean value and standard deviation identical to the observationally derived values as shown in the right panels of Figure~\ref{fig:line_ratio_evolution}) is attached to each object and the line ratio is corrected to $t = 220$ days by $r(220)~=~r(t)\times e^{\alpha (220 - t)}$ where $t$ is spectral phase in the unit of days. The distribution of the line ratios is then obtained for each Monte-Carlo simulation, and the KS test is performed for each simulated distribution. 

\begin{figure*}[!t]
\includegraphics[width = 12cm]{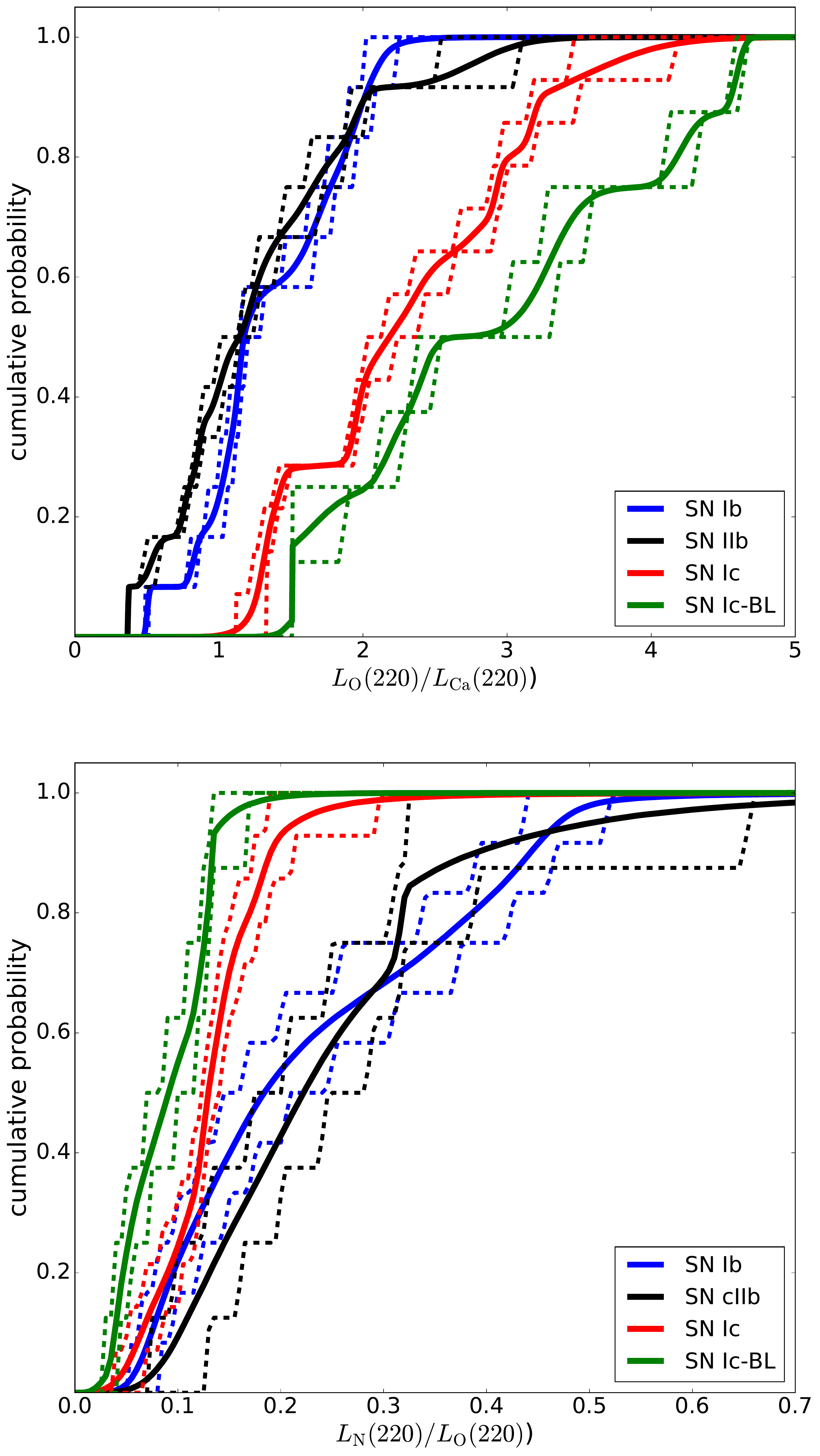}
\centering
\caption{The distribution of the line ratios of different SNe subtypes from 10$^4$ Monte-Carlo simulations. Solid lines are average values and dashed lines are 1$\sigma$ deviations. SNe of different subtypes are labeled by different colors.}
\label{fig:line_ratio_evolution_sim}
\vspace{4mm}
\end{figure*}

The results of the simulations are show in Figure~\ref{fig:line_ratio_evolution_sim}. The solid lines are the mean values of 10$^4$ simulations and the dashed lines are the 1$\sigma$ deviations. It is seen that even when the effect of time evolution is taken into consideration, the main conclusion is not affected. SNe Ib and SNe IIb have similar progenitor properties, with average and standard deviation of the KS test coefficients are 0.786$^{+ 0.205}_{- 0.353}$ for the [O I]/[Ca II] ratio and 0.565$^{+ 0.312}_{- 0.278}$ for the [N II]/[O I] ratio. SNe Ic have a systematically larger [O I]/[Ca II] ratio (therefore more massive progenitor) than SNe Ib, with the average and standard deviation of the KS test coefficient being 0.014$^{+ 0.007}_{- 0.008}$. The KS coefficient of different line ratios among SN subtypes are listed in Table~\ref{tab:KS_coefficient}. 
The effect of spectral evolution is therefore proven to be insignificant, which is expected since: (1) the [O I]/[Ca II] ratio evolves slowly; (2) although the evolution of [N II]/[O I] can be relatively fast, spectra from different subtypes in this work have similar phases, therefore the line evolution will not have any systematic effect on different subtypes.

\clearpage
\newpage
\begin{table}[!t]
\begin{threeparttable}
\caption{SN IIb in this work}
\label{tab:IIb_specList}
\setlength{\tabcolsep}{21pt}
\setlength{\extrarowheight}{5pt}
\begin{tabular}{|lcccc|}
\hline
SN name &date\tnote{a}&phase\tnote{b}&$E(B - V)\tnote{c}$&References\\
\hline
SN 1993J &1993/09/20&155&0.19&\cite{jerk15}, \citeS{matheson00}\\
&${\bf 1993/11/07}$&203&&\\
&1994/01/05&262&&\\
SN 1996cb&1997/06/03&150&0.03&\citeS{qiu99}\\
&${\bf 1997/07/01}$&178&&\\
&1997/12/05&335&&\\
SN 2001ig&${\bf 2002/10/08}$&274&0.10&\citeS{maund07}, \citeS{silverman09}\\
&2002/11/08&305&&\\
SN 2003bg&2003/08/20&153&0.02&\citeS{hamuy09}\\
&${\bf 2003/11/29}$&254&&\\
&2003/12/23&278&&\\
SN 2006T&${\bf 2006/11/26}$&288&0.08& \cite{maeda08}, \citeS{modjaz07}, \citeS{modjaz14}\\
&2006/12/25&317&&\\
SN 2008ax&2008/08/01&130&0.40&\citeS{modjaz14}, \citeS{tauben11}\\
&${\bf 2008/11/25}$&245&&\\
&2009/01/25&305&&\\
SN 2011dh&2011/12/18&182&0.07&\citeS{ergon15}, \citeS{ergon14}, \citeS{shivvers13}\\
&${\bf 2012/01/25}$&220&&\\
&2012/03/18&273&&\\
SN 2011fu&2012/02/22&133&0.10&\citeS{morales15}\\
&${\bf 2012/07/20}$&282&&\\
SN 2011hs&2012/03/18&160&0.17&\citeS{bufano14}\\
&${\bf 2012/06/21}$&211&&\\
&2012/10/22&334&&\\
SN 2012P&${\bf 2012/08/08}$&197&0.29&\citeS{fremling16}\\
(PTF 12os)&&&&\\
SN 2013ak&${\bf 2013/09/12}$&179&0.39\tnote{d}&\citeS{yaron12}\\
SN 2013df&2013/12/05&162&0.10&\citeS{maeda15}, \citeS{morales14}\\
&${\bf 2014/02/04}$&223&&\\
&2014/05/06&314&&\\\hline
    \end{tabular}
    \begin{tablenotes}
            \item{a}.~~ From Tables~\ref{tab:Ib_specList} to \ref{tab:IcBL_specList}, the spectra labeled by bold face is used for analyses in the main text.
            \item{b}.~~ Phase relative to the light curve maximum in most cases. However, the light curves are not available for all objects. The phase of an object, whose light curve is not available, relies on the literatures or IAUC reports by comparing thrie early spectra with those of reference SNe. This will introduce uncertainty in the phase determination. However, this will not have significant effect on the main conclusion of this work given the small uncertainty in the spectral phase determination as compared to the phase of the spectra used in this study.
            \item{c}.~~ From Tables~\ref{tab:Ib_specList} to \ref{tab:IcBL_specList}, objects labeled by ($+$) indicate the case where its extinction can not be calculated from light curve or Na I D absorption. The average $E(B - V)$ of SN Ib/c (0.36 mag) is adopted$^{\citeS{drout11}}$ for this case.
            \item{d}.~~ Calculated from Na I D absorption of spectra taken on 2013/04/01 ($t$ = +15 days)$^{\citeS{yaron12}}$.
  \end{tablenotes}
    \end{threeparttable}
\end{table}
\clearpage
\newpage
\begin{table}[!t]
\begin{threeparttable}
\caption{SN Ib in this work}
\label{tab:Ib_specList}
\setlength{\tabcolsep}{16pt}
\setlength{\extrarowheight}{5pt}
\begin{tabular}{|lcccc|}
\hline
SN name &date&phase&$E(B - V)$&References\\
\hline
SN 1990U&${\bf 1991/01/06}$&189&0.52&\citeS{modjaz14} \citeS{gomez94}, \citeS{matheson01}\\
SN 1990W&{\bf 1991/02/21}&180&0.36(+)&IAUC 5076, \citeS{tauben09}, \citeS{wheeler94}\\
&1991/04/21&239&&\\
SN 2004dk&${\bf 2005/05/11}$&263&0.34&\citeS{drout11}, \citeS{modjaz14}, \citeS{modjaz08}, \citeS{shivvers17}\\
&2005/07/09&322&&\\
SN 2004gq&${\bf 2005/08/26}$&249&0.25& \cite{maeda08}, \citeS{drout11}, \citeS{modjaz08}, \citeS{modjaz14}\\
&2005/10/08&292&&\\
SN 2004gv&2005-07-06&191&0.25\tnote{a}&\cite{maeda08}, \citeS{modjaz14}, \citeS{modjaz08}\\
&${\bf 2005/08/26}$&242&&\\
&2005/10/08&285&&\\
SN 2005bf&${\bf 2005/12/11}$&213&0.14&\citeS{modjaz14}, \citeS{folatelli06}\\
SN 2006F&${\bf 2006/06/30}$&175&0.54&CBET 368,  \cite{maeda08}, \citeS{drout11}\\
SN 2006gi&${\bf 2007/02/10}$&144&0.38&\citeS{tauben09}, \citeS{elmhamdi11}\\
SN 2007C&2007/05/17&121&0.64&\citeS{drout11}, \citeS{modjaz14}, \citeS{tauben09}\\
&${\bf 2007/06/20}$&155&&\\
SN 2007Y&${\bf 2007/09/22}$&200&0.11&\citeS{strit09}\\
&2007/11/30&269&&\\
SN 2009jf&${\bf 2010/06/19}$&245&0.12&\citeS{modjaz14}, \citeS{sahu11}, \citeS{valenti11}\\
&2010/07/08&264&&\\
iPTF13bnv&${\bf 2014/02/21}$&234&0.07&\citeS{fremling16}\\
&2014/05/28&330&&\\
\hline
    \end{tabular}
    \begin{tablenotes}
            \item{a}.~~ Calculated from Na I D absorption of spectra taken on 2005/01/11 ($t$ = +15 days)$^{\citeS{modjaz14}}$.
  \end{tablenotes}
    \end{threeparttable}
\end{table}
\clearpage
\newpage
\begin{table}[!t]
\begin{threeparttable}
\caption{SN Ic in this work}
\label{tab:Ic_specList}
\setlength{\tabcolsep}{17pt}
\setlength{\extrarowheight}{5pt}
\begin{tabular}{|lcccc|}
\hline
SN name&date&phase&$E(B - V)$&References\\
\hline
SN 1991N&${\bf 1992/01/09}$&277&0.12\tnote{a}&IAUC 5234, \cite{maeda08}, \citeS{matheson01}\\
SN 1994I&${\bf 1994/09/02}$&146&0.45& \citeS{modjaz14}, \citeS{filippenko95}, \citeS{rich96}\\
SN 1996aq&1997/02/11&160&0.36(+)&IAUC 6454, \citeS{tauben09}\\
&${\bf 1997/04/02}$&210&&\\
&1997/05/14&252&&\\
SN 1996D&${\bf 1996/09/10}$&199&0.36(+)&IAUC 6317, \citeS{tauben09}\\
SN 1997dq&1998/05/30&206&0.11\tnote{b}&IAUC 6770,\\
&${\bf 1998/06/18}$&225&&\citeS{modjaz14}, \citeS{matheson01}, \citeS{tauben09}\\
SN 2004aw&${\bf 2004/11/14}$&233&0.37&\citeS{modjaz14}, \citeS{tauben06}\\
SN 2004fe&${\bf 2005/07/06}$&249&0.32&\cite{maeda08}, \citeS{drout11}, \citeS{modjaz14}\\
&2005/08/26&300&&\\
SN 2004gk&${\bf 2005/07/10}$&223&0.47&\cite{maeda08}, \citeS{modjaz14}, \citeS{elmhamdi11}\\
SN 2004gt&${\bf 2005/05/24}$&152&0.10&\citeS{modjaz14}, \citeS{tauben09}, \citeS{galyam05}\\
SN 2005kl&${\bf 2006/06/30}$&213&0.29\tnote{c}&\cite{maeda08}, \citeS{drout11}, \citeS{modjaz14}\\
SN 2006ck&${\bf 2007/01/24}$&246&0.36(+)&IAUC 8713, \cite{maeda08}, \citeS{modjaz14}\\
SN 2011bm&${\bf 2011/12/17}$&227&0.06&\citeS{valenti12}\\
&2012/01/22&263&&\\
SN 2013ge&${\bf 2014/04/28}$&156&0.07&\citeS{drout16}\\
&2014/10/23&334&&\\
\hline
    \end{tabular}
    \begin{tablenotes}
            \item{a}. ~~ Calculated from Na I D absorption of spectra taken on 1991/04/07 ($t$ = +0 days)$^{\citeS{matheson01}}$.
            \item{b}. ~~ Calculated from Na I D absorption of spectra taken on 1997/11/08 ($t$ = +3 days)$^{\citeS{matheson01}}$.
            \item{c}. ~~ Calculated from Na I D absorption of spectra taken on 2005/11/25 ($t$ = -4 days)$^{\citeS{modjaz14}}$.
         
  \end{tablenotes}
    \end{threeparttable}
\end{table}

\clearpage
\newpage
\begin{table}[!t]
\begin{threeparttable}
\caption{SN Ic-BL in this work}
\label{tab:IcBL_specList}
\setlength{\tabcolsep}{16pt}
\setlength{\extrarowheight}{5pt}
\begin{tabular}{|lcccc|}
\hline
SN name &date&phase\tnote{a}&$E(B - V)$&References\\
\hline
SN 1997ef&${\bf 1998/09/21}$&282&0.00&\citeS{modjaz14}, \citeS{matheson01}, \citeS{iwamoto00}, \citeS{mazzali00}\\
SN 1998bw&${\bf 1998/11/26}$&198&0.06&\citeS{clocchiatti11}, \citeS{modjaz16}, \citeS{patat01}\\
&1999/04/12&335&&\\
SN 2002ap&2002/08/09&183&0.08&\citeS{modjaz16}, \citeS{foley03}, \citeS{yoshii03}\\
&${\bf 2002/10/01}$&236&&\\
&2002/11/06&272&&\\
SN 2005kz&2005/06/30&215&0.46&IAUC 8639, \cite{maeda08}, \citeS{drout11}\\
SN 2005nb&2006/06/30&183&0.36(+)&IAUC 8657, \cite{maeda08}, \citeS{modjaz14}\\
SN 2006aj&2006/09/21&206&0.15&\citeS{modjaz14}, \citeS{modjaz16}, \citeS{modjaz06}\\
SN 2007I&2007/07/15&195&0.36(+)&CBET 808, \citeS{modjaz14}, \citeS{tauben09}\\
SN 2012ap&2012/09/21&272&0.45&\citeS{milli15}\\
&2012/11/16&328&&\\
\hline
    \end{tabular}
    \end{threeparttable}
\end{table}

\clearpage
\newpage
\begin{table}[!t]
  \begin{threeparttable}
\caption{Early parameters}
\label{tab:early_para}
\centering
\setlength{\extrarowheight}{3pt}
\setlength{\tabcolsep}{19pt}
\begin{tabular}{|lcccc|}
\hline
SN name &type\tnote{a}&$\Delta m_{15}$&$v_{\rm ph}$ (km s$^{-1}$)\tnote{b}&references\\\hline
SN 1993J&IIb&0.971$^{+0.052}_{-0.078}$&8000$^{+1000}_{-1000}$&\cite{lyman16}, \citeS{matheson00}\\
SN 1994I&Ic&1.745$^{+0.091}_{-0.140}$&11500$^{+1000}_{-1400}$&\cite{lyman16}, \citeS{rich96}\\
SN 1996cb&IIb&0.764$^{+0.059}_{-0.072}$&8500$^{+1300}_{-1000}$&\cite{lyman16}, \citeS{qiu99}\\
SN 1997ef&Ic-BL&0.331$^{+0.025}_{-0.050}$&13000$^{+5000}_{-5000}$&\citeS{iwamoto00}\\
SN 1998bw&Ic-BL&0.782$^{+0.065}_{-0.079}$&19500$^{+1700}_{-1000}$&  \cite{lyman16}, \citeS{clocchiatti11}\\
SN 2002ap&Ic-BL&0.983$^{+0.045}_{-0.071}$&13000$^{+2000}_{-1000}$&\cite{lyman16}, \citeS{foley03}\\
SN 2003bg&IIb&0.520$^{+0.101}_{-0.122}$&8000$^{+1000}_{-1000}$&\cite{lyman16}, \citeS{hamuy09}\\
SN 2004aw&Ic&0.564$^{+0.047}_{-0.056}$&11000$^{+1000}_{-1900}$&\cite{lyman16}, \citeS{tauben06}\\
SN 2004dk&Ib&0.468$^{+0.041}_{-0.049}$&9200$^{+1400}_{-1000}$&\cite{lyman16}, \citeS{drout11}, \citeS{haru08}\\
SN 2004fe&Ic&1.110$^{+0.078}_{-0.100}$&11000$^{+1000}_{-1000}$&\cite{lyman16}, \citeS{strit18}\\
SN 2004gq&Ib&0.876$^{+0.066}_{-0.082}$&13000$^{+1000}_{-1500}$&\cite{lyman16}, \citeS{drout11}\\
SN 2004gt&Ic&0.634$^{+0.058}_{-0.068}$&10400$^{+1200}_{-1200}$&\citeS{strit18}\\
SN 2004gv&Ib&0.777$^{+0.058}_{-0.072}$&9900$^{+1400}_{-1400}$&\citeS{strit18}\\
SN 2005bf&Ib&0.433$^{+0.057}_{-0.058}$&7500$^{+1800}_{-1000}$&\cite{lyman16}, \citeS{folatelli06}\\
SN 2005kl&Ic&0.835$^{+0.087}_{-0.095}$&10400$^{+1200}_{-1200}$&\citeS{bianco14}\\
SN 2006aj&Ic-BL&1.367$^{+0.095}_{-0.127}$&18000$^{+1000}_{-1000}$&\cite{lyman16}, \citeS{bianco14}\\
SN 2006T&IIb&0.951$^{+0.071}_{-0.090}$&7500$^{+1000}_{-1000}$&\cite{lyman16}, \citeS{strit18}\\
SN 2007C&Ib&0.966$^{+0.062}_{-0.084}$&11000$^{+1000}_{-1400}$&\cite{lyman16}, \citeS{drout11}\\
SN 2007Y&Ib&1.012$^{+0.052}_{-0.080}$&9000$^{+1000}_{-1700}$&\cite{lyman16}, \citeS{strit09}\\
SN 2008ax&Ib&1.041$^{+0.061}_{-0.087}$&7500$^{+2100}_{-1000}$&\cite{lyman16}, \citeS{tauben11},  \citeS{tsve09}\\
SN 2009jf&Ib&0.476$^{+0.080}_{-0.026}$&9500$^{+2100}_{-1000}$&\cite{lyman16}, \citeS{sahu11}\\
SN 2011bm&Ic&0.172$^{+0.016}_{-0.018}$&9000$^{+1000}_{-1000}$&\cite{lyman16}, \citeS{valenti12}\\
SN 2011dh&IIb&0.923$^{+0.071}_{-0.088}$&7000$^{+1000}_{-1000}$&\cite{lyman16}, \citeS{ergon14}\\
SN2011fu&IIb&0.816$^{+0.039}_{-0.063}$&8300$^{+1750}_{-1750}$&\citeS{morales15}\\
SN 2011hs&IIb&1.252$^{+0.067}_{-0.101}$&8000$^{+1000}_{-1000}$&\cite{lyman16}, \citeS{bufano14}\\
SN 2012P\tnote{c}&IIb&0.786$^{+0.073}_{-0.084}$&7500$^{+1750}_{-1750}$&\citeS{fremling16}\\
(PTF12os)&&&&\\
SN 2012ap&Ic-BL&0.908$^{+0.070}_{-0.087}$&19100$^{+6000}_{-6000}$&\citeS{milli15}\\
SN 2013ak&IIb&0.961$^{+0.081}_{-0.099}$&8000$^{+1000}_{-1000}$&\citeS{brown14}\\
SN 2013df&IIb&1.059$^{+0.035}_{-0.052}$&8450$^{+1000}_{-1000}$&\citeS{morales14}, \citeS{szalai16}\\
SN 2013ge&Ic&0.732$^{+0.061}_{-0.074}$&10400$^{+2200}_{-2200}$&\citeS{drout16}\\
iPTF13bvn&Ib&1.170$^{+0.069}_{-0.088}$&8000$^{+1000}_{-1000}$&\cite{lyman16}, \citeS{fremling16}\\
\hline
\end{tabular}
\begin{tablenotes}
            \item{a}.~~ See Tables \ref{tab:Ib_specList} to \ref{tab:IcBL_specList} for references.
            \item{b}.~~ Object with marker (*) means that its photometric velocity is taken as the average value of SN subtype reported in Lyman et al. 2016$^{\cite{lyman16}}$.
            \item{c}.~~ PTF12os does not have a reported $V-$band light curve. The absolute flux scales of early phase spectra are anchored by $r-$band photometric data$^{\citeS{fremling16}}$ and $V-$band light curve is calculated from the calibrated early phase spectra. 
  \end{tablenotes}
  \end{threeparttable}
\end{table}

\clearpage
\newpage
\begin{table}[!t]
\begin{threeparttable}
\caption{KS coefficients between different SN subtypes after phase is corrected}
\label{tab:KS_coefficient}
\setlength{\tabcolsep}{11pt}
\setlength{\extrarowheight}{8pt}
\begin{tabular}{|lcccccc|}
\hline
Group&Ib/IIb&Ib/Ic&Ib/Ic-BL&IIb/Ic&IIb/Ic-BL&Ic/Ic-BL\\\hline
$[$N II$]$/$[$O I$]$\tnote{a}&0.565$^{+0.311}_{-0.278}$&0.052$^{+0.187}_{-0.038}$&0.014$^{+0.031}_{-0.010}$&0.020$^{+0.085}_{-0.018}$&0.010$^{+0.040}_{-0.008}$&0.106$^{+0.274}_{-0.072}$\\
$[$O I$]$/$[$Ca II$]$&0.786$^{+0.205}_{-0.353}$&$0.014^{+0.007}_{-0.008}$&0.004$^{+0.010}_{-0}$&$0.011^{+0.014}_{-0.005}$&$0.014^{+0}_{-0.010}$&0.441$^{+0.209}_{-0.211}$\\
\hline
    \end{tabular}
    \begin{tablenotes}
            \item{a}. ~~ When comparing [N II]/[O I], SNe eIIb are excluded from the SNe IIb sample, since their `[N II]' flux may be contaminated by H$\alpha$. For [O I]/[Ca II], SNe eIIb are included in the KS test.
         
  \end{tablenotes}
    \end{threeparttable}
\end{table}

\clearpage
\newpage

\vspace{1cm}
\noindent
{\huge Acknowledgements}\\
QF acknowledges support by MEXT scholarship awarded by Ministry of Education, Culture, Sports, Science and Technology, Japan. KM acknowledges support by JSPS KAKENHI Grant (18H04585, 18H05223, 17H02864). AGY is supported by the EU via ERC grant No. 725161, the ISF, the BSF Transformative program and by a Kimmel award.

\vspace{1cm}
\noindent
{\huge Author Contributions}\\
QF led the nebular spectral analysis and the manuscript preparatoin. KM contributed to the initiation of the project together with QF. KM and AG organized the efforts for interpretation of the results and assisted in manuscript preparation. HK contributed to the spectral analysis and interpretations. FS assisted the early and nebular data analysis. All authors contributed to discussion. 

\vspace{1cm}
\noindent
{\huge Correspondance}\\
Correspondence and requests for materials should be addressed to Q. Fang (fangql@kusastro.kyoto-u.ac.jp) and K. Maeda (keiichi.maeda@kusastro.kyoto-u.ac.jp). 

\vspace{1cm}
\noindent
{\huge Competing Interests}\\
The authors declare that they have no competing financial interests.

\end{document}